# A high-resolution molecular spin-photon interface at telecommunications wavelengths

**Authors:** Leah R. Weiss[1]†, Grant T. Smith[1]†, Ryan A. Murphy[2,3]†, Bahman Golesorkhi[2], José A. Méndez Méndez [1], Priya Patel[2], Jens Niklas[4], Oleg G. Poluektov[4], Jeffrey R. Long[2,3,5]*, David D. Awschalom[1,6,7]*

**Affiliations:**

[1]Pritzker School of Molecular Engineering, University of Chicago, Chicago, IL, 60637, USA.

[2]Department of Chemistry, University of California, Berkeley, CA 94720, USA.

[3]Materials Sciences Division, Lawrence Berkeley National Laboratory, Berkeley, CA 94720, USA.

[4]Chemical Sciences and Engineering Division, Argonne National Laboratory, Lemont, Illinois 60439, USA.

[5]Department of Chemical and Biomolecular Engineering and Department of Materials Science and Engineering, University of California, Berkeley, CA 94720, USA.

[6]Department of Physics, University of Chicago, Chicago, IL 60637, USA.

[7]Center for Molecular Engineering and Materials Science Division, Argonne National Laboratory, Lemont, IL 60439, USA.

*Corresponding authors. Email: awsch@uchicago.edu (D.D.A.), jrlong@berkeley.edu (J.R.L.)

†These authors contributed equally.

**Abstract:**
Optically addressable electronic spins in polyatomic molecules are a promising platform for quantum information science with the potential to enable scalable qubit design and integration through atomistic tunability and nanoscale localization. However, optical state- and site-selection are an open challenge. Here we introduce an organo-erbium spin qubit in which narrow (MHz-scale) optical and spin transitions couple to provide high-resolution access to spin degrees of freedom with telecommunications frequency light. This spin-photon interface enables demonstration of optical spin polarization and readout that distinguishes between spin states and magnetically inequivalent sites in a molecular crystal. Operation at frequencies compatible with mature photonic and microwave devices opens a path for engineering scalable, integrated molecular spin-optical quantum technologies.

**Main Text:**

Probing and controlling increasingly complex states of matter is a frontier of quantum science. The spin states of polyatomic molecules can form coherent two-level systems (qubits): modular building blocks for quantum technologies constructed at the atomic level. This approach to engineering quantum systems at scale leverages the tools of synthetic chemistry for atomistic design, nanoscale localization, and multi-component assembly (*1–3*). These molecular materials could further be integrated in non-standard environments, from the surface of a solid-state device to the interior of a cell. However, interfacing this nascent class of qubits with light—a promising approach to scalable readout, control, and coupling (*4-9*) — remains in its infancy. For applications from quantum sensing to networking (*10–13*), a spin-photon interface must distinguish both between qubit states (to enable high-fidelity initialization, readout, and light-matter interactions) and distinct physical sites (to enable multi-qubit architectures).

To address these requirements, we introduce an optically addressable molecular spin qubit based on an erbium metal-organic complex—a representative of the larger class of rare-earth ions that can feature magnetic ground states and shielded optical transitions between 4f-orbitals (*14*). This combination enables a high-resolution electronic spin-photon interface. We demonstrate how all-optical spin spectroscopy can resolve magnetic interactions, distinguish between magnetically inequivalent sites, and confirm optical spin pumping and initialization of the coherently controllable ground state of a molecular ensemble. We highlight operation with erbium to take advantage of optical transitions at telecommunications wavelengths. This wavelength regime allows access to optical components developed at-scale for telecommunications applications and minimizes losses in long-distance photon transport over optical fibers for quantum networking (*15*). It further sets the operating frequency in the transparency window of silicon (*16, 17*) for compatibility with state-of-the-art integrated photonic technology required for scalable, on-chip spin-optical interconnects and molecular quantum optics (*5, 18–25*).

This approach builds on a deep foundation of work in two domains where the unique properties of 4f-electrons are enabling. In solid-state quantum optics, the atom-like optical transitions of rare-earth ions doped in covalent solids have found wide-ranging applications from the construction of lasers and amplifiers to early building blocks of quantum networks and transducers (*13, 14, 26, 27*). Recent demonstrations of narrow optical linewidths in rare-earth molecules suggest that these properties can further be ported to molecular architectures (*28–31*). In parallel, the large, anisotropic magnetic moments of lanthanides underly demonstrations of the power of synthetic chemistry to design the spin properties of molecular materials, as exemplified in the engineering of single-molecule magnets, magnetic resonance contrast agents, and molecular qubits (*32–38*).

Here, as a proof-of-principle for an optically addressable molecular lanthanide qubit, we study Cs[Er(hexafluoroacetylacetonate)$_4$] (denoted Er(hfa)$_4$) diluted within mm-scale single crystals of isostructural Y(hfa)$_4$ (*39, 40*). The compound (Fig. 1D) is air-stable, amenable to thermal evaporation for device integration, and crystallizes via ionic and van der Waals interactions in chains of alternating organo-lanthanide and cesium counter-ions to provide net charge neutrality in the solid state (Fig. 1B) (*39*). As shown schematically in Fig. 1A, optical transitions of the central erbium ion occur at ~1.5 μm between the total angular momentum $J$ = 15/2 ground state and $J$ = 13/2 excited state. These manifolds are split into mixed-$M_J$ sublevels via the crystal-field interaction. In the low-symmetry ligand environment probed here, the THz-scale splitting of the crystal-field levels within the ground- and excited-state manifolds allows the transitions between the lowest two levels to be spectrally isolated. As shown in Fig. 1E, we measured photoluminescence as a function of laser excitation wavelength (photoluminescence excitation, PLE) in a closed-cycle helium-4 cryostat at 3.4 K (Methods, Fig. 1C schematic) and observed transitions from the ground state to a series of three excited-state crystal field levels. The lowest-energy transition features an inhomogeneous full-width-half-max (FWHM) linewidth of 945(8) MHz (inset of Fig. 1E) with a spatial average across a bulk crystal giving 3.7 GHz (fig. S2). This inhomogeneous broadening is comparable to that observed for erbium doped within high-quality solid-state crystal lattices (*41–43*). Note that for optical spectroscopy we use a perdeuterated sample (fully deuterated erbium compound and yttrium host) to enhance luminescence intensity and reduce quenching by C-H vibrations (*43*), resulting in an 8.66(1) μs optical lifetime (fig. S3).

To demonstrate microwave addressability of the ground state, we performed electron spin resonance (ESR) spectroscopy. In the presence of an external magnetic field, the Zeeman interaction lifts the two-fold degeneracy of the ground-state (Fig. 1A) and we can address the resulting microwave transition. As shown in Fig. 2A, the continuous-wave (cw) powder ESR spectrum at 10 K reflects a spin-1/2 with an anisotropic *g*-factor far from the free electron g-factor of 2.0023 ($g_x < g_y = 7.45$, $g_z = 10.76$) (*45*). As shown in Fig. 2B, the echo-detected field-swept (EDFS) ESR spectrum in a bulk single crystal (with microwave field approximately aligned along the *a*-axis of the crystal) shows satellite lines consistent with the nuclear spin $I = 7/2$ of the 23% abundant $^{167}$Er isotope (the only stable Er isotope with a nuclear spin).

Next, we probed the coherence of the ground-state spin transition using pulsed ESR techniques. We detected the intensity of the Hahn echo signal as a function of pulse length at the peak of the EDFS spectrum (270 mT) (*46*). Coherent Rabi oscillations were observed (Fig. 2C, pulse sequence shown below experimental data) with an exponential decay time of 80-100 ns (shaded exponential decay region), and the expected square-root dependence of the Rabi rate on applied microwave power was confirmed (Fig. 2D). We then used a Hahn echo decay sequence, varying the free evolution time τ between π/2-pulse and π-pulse (inset Fig. 2E), to observe echo decay over the same timescale as the Rabi oscillations (100(2) ns). The echo decay curve features oscillations corresponding to electron-spin-echo envelope modulation (ESEEM) due to hyperfine interactions (*47*). As shown in fig. S4, the observed ESEEM frequency is consistent with the free precession frequencies of both fluorine and hydrogen as expected from the presence of both atoms on the ligands (Larmor frequencies of $^{19}$F and $^1$H are indistinguishable at these magnetic fields due to similar gyromagnetic ratios). To assess the extent to which the loss of coherence is limited by spin-lattice relaxation we performed an inversion recovery sequence (Fig. 2F inset) which yields a spin polarization recovery time of 198(2) ns (Fig. 2F) and suggests that coherence decay is not limited by interactions with phonons but may rather be dominated by field fluctuations from nearby electron and nuclear spins in the crystal (*48, 49*). The echo signal is not measurable at higher temperatures, consistent with the exponential increase in continuous-wave ESR linewidth as a function of temperature (fig. S5) expected from Orbach relaxation due to phonon-mediated mixing with higher lying crystal-field states (*50*).

Strong spin-orbit coupling not only yields the large, anisotropic ground-state *g*-factor observed via ESR, but can also provide a difference in *g*-factor between the ground and excited states. Here we utilize this feature in combination with narrow optical and spin linewidths to demonstrate a direct, ensemble-resolved spin-optical interface. Under an applied magnetic field, Zeeman splitting is optically resolved across the ensemble, as shown in Fig. 3 where we plot PLE spectra as a function of magnetic field. As shown schematically in the stick spectrum and energy level diagram in Fig. 3B, the origin of the four peaks can be explained with a difference in effective g-factor between ground and excited state ($g_e$ and $g_g$), which produces a set of approximately spin-conserving transitions separated by $|g_e - g_g|\mu_B B/h$ and approximately spin-non-conserving peaks separated by $|g_e + g_g|\mu_B B/h$, where $B$ is the strength of the external magnetic field, $\mu_B$ is the Bohr magneton, and $h$ is the Planck constant. We use the difference in observed photoluminescence intensity with cryostat thermometers reading ~4 K and ~60 mK to assign transitions A and C to transitions from the lower ground state ($|\downarrow_g\rangle$ to $|\downarrow_e\rangle$ and $|\uparrow_e\rangle$ respectively) and D and B to transitions from the higher ground-state ($|\uparrow_g\rangle$ to $|\downarrow_e\rangle$ and $|\uparrow_e\rangle$, respectively), as labelled on the spectrum in Fig. 3A and on the level diagram in Fig. 3B. These assignments allow us to quantify optically the *g*-factors in the ground and excited states from measured PLE as a function of magnetic field (Fig. 3C) and extract values of $g_g$ = 10.8(1) and $g_e$ = 12.9(1) from a linear fit to the measured splitting at this crystal orientation (Fig. 3D). The optically measured value of 10.8(1) is in good agreement with the approximate $g_z$ value inferred from the lowest peak in the powder ESR spectrum, confirming robust coupling of the optical transitions to ground- and excited-state spin degrees of freedom and opening opportunities for spin-optical coupling of collective spin states of the ensemble (*26, 51*).

Underlying the inhomogeneity that broadens ensemble spectra, the optical transitions of individual molecules possess much narrower lines, here accessed via a two-tone spectral hole-burning technique. Differential photoluminescence is detected as a function of detuning between two narrow-line lasers (pump and probe). The pump, centered on the spin-conserving transitions, shelves a portion of the addressed sub-ensemble population in the excited state and the probe is swept to reveal the spectral hole where ground-state population is depleted. The observed linewidth of this central hole is 10.9(5) MHz, representing an upper bound on the homogenous linewidth of the optical transition under these conditions (Fig. 4B). Further, due to relaxation between spin sublevels in both the ground and excited states at these temperatures (3.2-4 K), spectral holes are also visible at detuning corresponding to both the ground- and excited-state spin splitting, as well as their difference (shown schematically in Fig. 4A and experimentally in Fig. 4C). In this way, we are able to access the molecular spin structure at low magnetic fields with a resolution well below the inhomogeneous linewidth.

At sufficiently high fields (24 mT shown here), spectra of these ground- and excited-state spin transitions reveal an additional splitting consistent with the presence of magnetically distinct sites, as expected from the crystal structure for slight misalignment of the magnetic field from the *a*-axis (Fig. 4D). To verify this site-resolution, we measure the splitting between transitions C and D as a function of the angle of the external magnetic field in the plane perpendicular to the optical axis (full PLE data as a function of angle are provided in fig. S6). The data (black circles, Fig. 4F) can be modelled by two magnetically inequivalent sites with *g*-tensors related by rotation (blue and purple solid lines), as suggested by the crystal structure. The full optical resolution of spin states and lambda level structure demonstrated here opens an opportunity to control spin polarization through optical pumping but requires that spin relaxation be slower than optical cycling.

To demonstrate this capacity for optical spin initialization, we performed optical spectroscopy in a dilution refrigerator at sub-Kelvin temperatures (Methods) where spin-lattice relaxation is minimized. As shown in fig. S7, we used the spin-dependent ensemble optical transitions described above to directly probe the thermal occupation of each spin sub-level and confirm that the bulk crystalline sample is thermalized with an approximate spin temperature of ~600 mK under illumination, as determined through optical Boltzmann thermometry. A time-dependent loss of photoluminescence under pulsed excitation (Fig. 5B) was observed, consistent with laser pumping to a dark state (schematic shown in Fig. 5A) that occurs at sub-Kelvin temperatures (blue) and is not apparent at ~4 K (grey), as expected from the fast relaxation at higher temperatures measured via ESR. To identify the state into which pumping occurs, we again performed two-tone spin spectroscopy with an electro-optic modulator to generate sidebands detuned from the carrier by an applied microwave frequency (Methods) and detected the differential photoluminescence (holes and anti-holes) that emerge relative to a Lorentzian background. As shown experimentally in Fig. 5C and D and schematically in Fig. 5A, the pattern of positive and negative peaks is consistent with laser pumping into the opposite electronic spin state. These results highlight the power of this direct, high-resolution spin-optical interface to enable detection and preparation of the ground-state spin population both under thermal initialization and selective spin-pumping.

The demonstrated link between the optical and microwave regimes within a nanoscale molecular system represents a key step toward developing more complex all-optical control of molecular qubits and engineering hybrid quantum systems with interactions mediated by synthetically tunable molecules. In future work the mechanisms that control both inhomogeneous and homogeneous optical linewidths should be explored further (*52*, *53*) to enable optimization by synthetic design. In the spin domain, the experimental evidence for fast relaxation at He-4 temperatures suggests that the design rules explored for single-molecule magnets may enable higher temperature spin initialization through ligand modifications (*35*). These all-optical spin spectroscopy techniques may also find utility in broader contexts for the study of molecular lanthanide spins with high spatial resolution afforded by confocal microscopy and with access to excited-state spin dynamics currently out of reach (*34*, *35*). Future work may also take advantage of localized, high-quality optical transitions for integration in silicon nanophotonics, not only to explore cavity quantum electrodynamics but also as a promising pathway toward the control of single spin-bearing molecules (*20*, *55*, *56*) and synthetically assembled arrays (*57*). The demonstrated capacity for optically resolving and controlling molecular magnetic states thereby opens new avenues both for engineering molecular quantum technologies and new tools for understanding molecular magneto-optics.

**Acknowledgments:** We thank S. L. Bayliss and Y. Tsaturyan for helpful discussions.

**Funding:** This work was supported by Q-NEXT, a U.S. Department of Energy Office of Science National Quantum Information Science Research Centers and by the U.S. Department of Energy, Office of Science, Basic Energy Sciences under awards DE-SC0019356 and DE-SC0025176. Preliminary synthesis and characterization was funded by the U.S. Department of Energy, Office of Science, Office of Basic Energy Sciences, Division of Chemical Sciences, Geosciences, and Biosciences at LBNL under contract no. DE-AC02-05CH11231. We thank the Quantum Information Science and Engineering Network for partial support of R.A.M. through NSF award DMR-1747426. L. R. W. acknowledges support from the University of Chicago/Advanced Institute for Materials Research Joint Research Center. The ESR work was supported by the U.S. Department of Energy, Office of Science, Office of Basic Energy Sciences, Division of Chemical Sciences, Geosciences, and Biosciences, through Argonne National Laboratory under Contract No. DE-AC02-06CH11357.

**Author contributions:** R.A.M. prepared and characterized the materials with assistance from B.G. and P.P. L.R.W. and G.T.S. performed spin-optical experiments and analysis with assistance from J.M.M. J.N. and O.G.P. assisted with ESR measurements. All authors discussed the results and contributed to manuscript preparation. J.R.L. and D.D.A. advised on all efforts.

**Competing interests:** Authors declare that they have no competing interests.

**Data and materials availability:** The data underlying this study will be made available at Zenodo. Requests for materials should be directed to D.D.A and J.R.L.

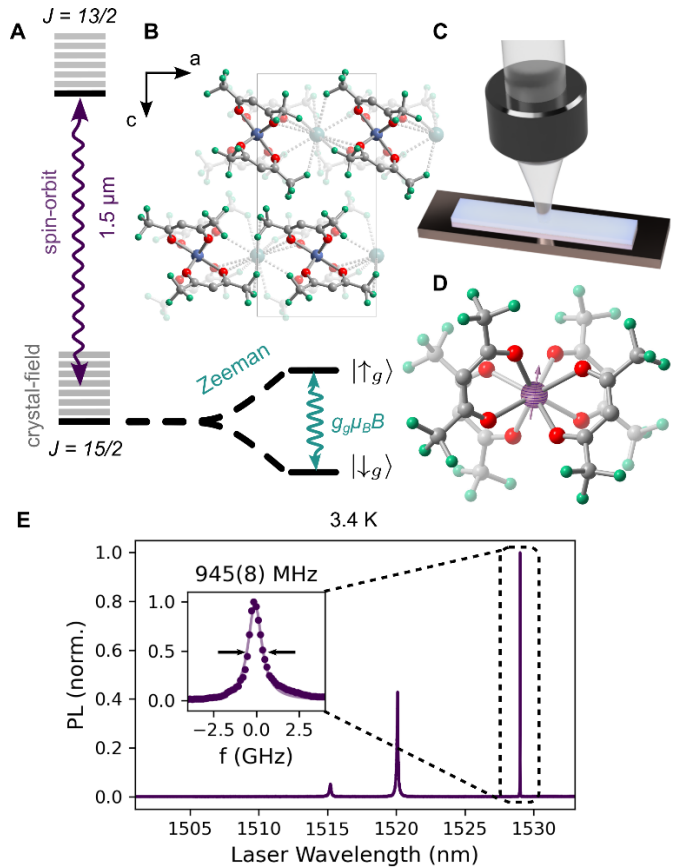

**Fig. 1. Optically addressable molecular erbium.** (**A**) Level-structure diagram for trivalent erbium with a 1.5 μm optical transition between spin-orbit coupled ground-state with total angular momentum $J = 15/2$ and excited-state with $J = 13/2$. Degeneracy within the ground- and excited-state manifolds is lifted by crystal-field interactions (shown in grey) and ground-state Zeeman splitting (green) in an external magnetic field, which sets the qubit energy splitting $g_g\mu_B B$ with $g_g$ the ground-state gyromagnetic ratio, $\mu_B$ the Bohr magneton, and $B$ the strength of the static external magnetic field. (**B**) Packing in unit cell of host Y(hfa)$_4$ and isostructural dopant Er(hfa)$_4$ (**D**). Purple, red, grey, green, turquoise, and blue spheres represent erbium, oxygen, carbon, fluorine, cesium, and yttrium atoms, respectively. Hydrogen atoms have been removed for clarity. (**B**) Experimental setup with bulk molecular crystal mounted on a copper in a helium-4 cryostat or dilution refrigerator with optical excitation and collection via microscope objective (Methods). (**E**) Photoluminescence (PL) as a function of laser wavelength measured at 3.4 K in d$_4$-Er(hfa)$_4$ doped in d$_4$-Y(hfa)$_4$ (200 ppm) with minimum observed FWHM inhomogeneous linewidth $\Gamma_{inh} = 945(8)$ MHz (inset).

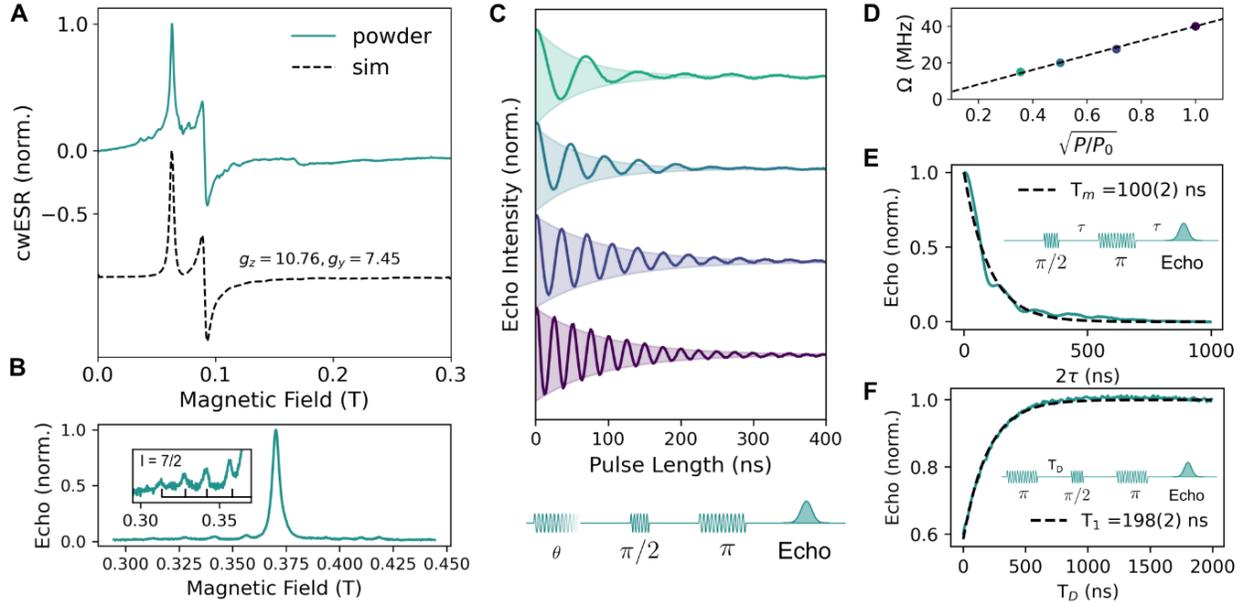

**Fig. 2. Microwave addressable coherent ground-state.** (**A**) Continuous-wave (cw) ESR of Er(hfa)$_4$ powder (shown in green at 9.5 GHz, 200 ppm doping in the Y analog, 10 K) and corresponding simulation (black dashes) reproducing turning points with principal components of the anisotropic g-tensor ($g_x < g_y = 7.45$, $g_z = 10.76$). (**B**) Field-swept, echo-detected ESR on a bulk crystal (200 ppm Er doping) at 3.7 K exhibiting fine-structure (zoomed inset) consistent with hyperfine splitting of the $^{167}$Er, a spin $I = 7/2$ nucleus. (**C**) Coherent Rabi oscillations measured at a field position corresponding to peak echo intensity in (B) as a function of microwave power with pulse sequence denoted below with variable length of the first pulse (θ). (**D**) Square-root dependence of oscillation frequency on relative microwave power $P/P_0$ with dashed line $\propto \sqrt{P/P_0}$ and $P_0$ maximum applied microwave power. (**E**) Hahn echo decay as a function of $2\tau$ and fitted exponential decay (dashed line least-squares fit to exponential decay with phase memory time $T_m$ = 100(2) ns). (**F**) Inversion recovery sequence (inset) with measured echo-detected polarization as a function of delay time $T_D$ following inversion pulse (green) and fitted exponential echo recovery timescale (dashed line, $T_1$ = 198(2) ns).

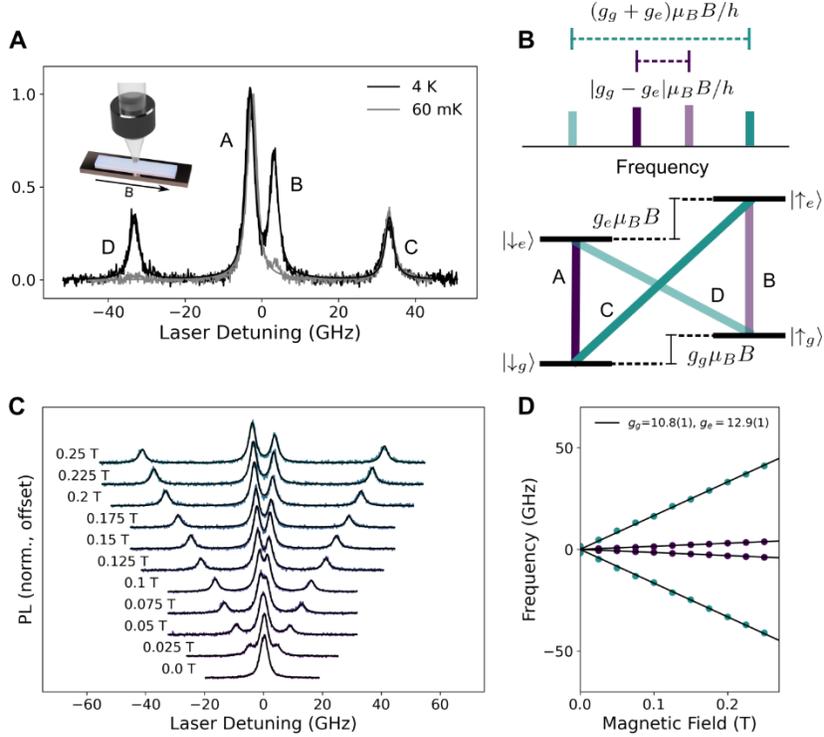

**Fig. 3. Ensemble-resolved spin-optical interface.** (**A**) Photoluminescence as a function of laser excitation frequency (PLE spectra) with a 200 mT field applied along the long axis of a bulk crystal of $d_4$-Er(hfa)$_4$ doped in $d_4$-Y(hfa)$_4$ (50 ppm) at $T \sim 4$ K (black) and $T \sim 60$ mK (grey), enabling assignment of transitions from the lower ($|\downarrow_g\rangle$) and upper ($|\uparrow_g\rangle$) ground-state sub-levels. (**B**) Level diagram for optical transitions between spin states and corresponding stick spectrum labelled to correspond to the observed transitions in (A). (**C**) Dependence of PLE spectra on applied magnetic field and with solid lines corresponding to least-squares fit to four Lorentzian peaks used to extract (**D**) transition frequencies as a function of applied magnetic field and corresponding g-factors for ground- and excited-state spins $g_g = 10.8(1)$ and $g_e = 12.9(1)$, respectively.

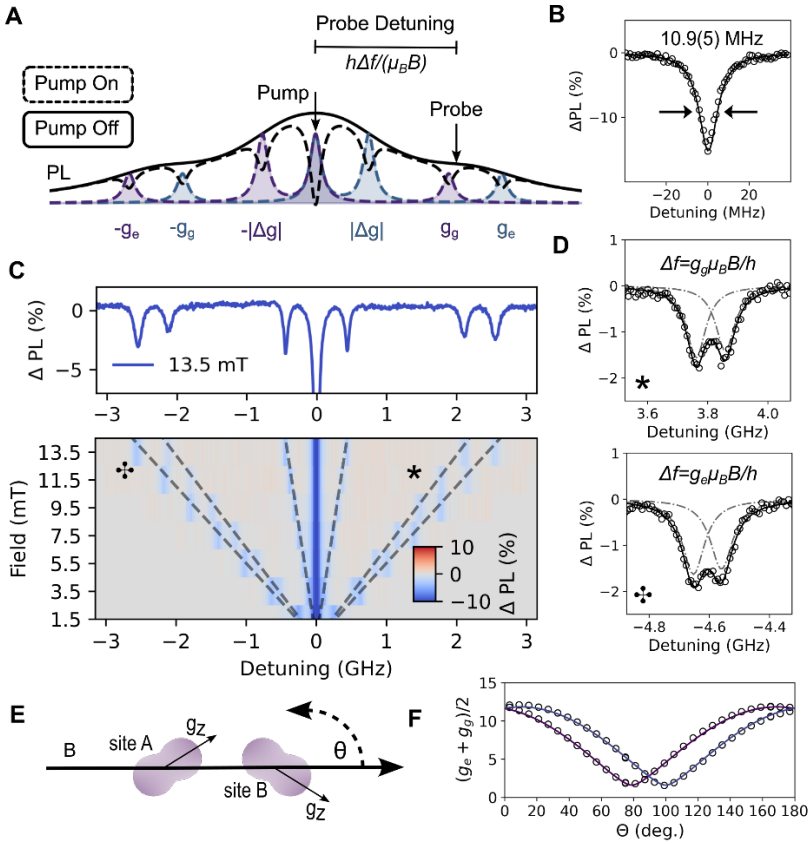

**Fig. 4. Molecular spin and site distinguishability with sub-GHz resolution**. (**A**) Schematic two-color spectral hole-burning experiment in regime of fast relaxation between ground-state spin populations in which a pump tone is applied to produce a central hole with a width corresponding to the homogeneous sub-ensemble linewidth ($\Gamma_{\text{hom}}$) and side-holes detuned by $\pm|\Delta f|$ equal to (1) $|\Delta g|\mu_B B/h$, (2) $g_g \mu_B B/h$, (3) $g_e \mu_B B/h$. Note that the pattern assumes resonance with two sub-ensembles with pump resonant with transition A (blue) or B (purple) with $\frac{\Delta g \mu_B B}{h} < \Gamma_{\text{inh}}$. (**B**) Central spectral hole width measured at 3.2 K and bulk crystal (200ppm erbium doping) using two distinct lasers to control pump and probe frequencies (Methods) with Lorentzian FWHM linewidth $\Gamma_{\text{hom}} = 10.9(5)$ MHz (solid line, least squares fit) at $B = 24$ mT. (**C**) Hole-burning spectrum as a function of applied magnetic field showing transitions and dashed lines corresponding to (1), (2), (3) defined above with $g_e$ and $g_g$ values measured in Fig. 3. (**D**) Further splitting of transition 2 (upper, marked ⋆) and 3 (lower, marked ✤) corresponding to ground- and excited-state splitting at $B = 24$ mT. Both resonances show a further splitting consistent with the presence of two magnetically inequivalent molecular sites (dashed grey lines). (**E**) Schematic of two g-tensors corresponding to two molecular sites related by rotation. (**F**) Dependence of the spectral splitting observed between transitions C and D yields an average g-value of $(g_g + g_e)/2$ as a function of the angle $\theta$ of the applied magnetic field in the lab frame consistent with two anisotropic g-tensors related by rotation (blue and purple solid lines).

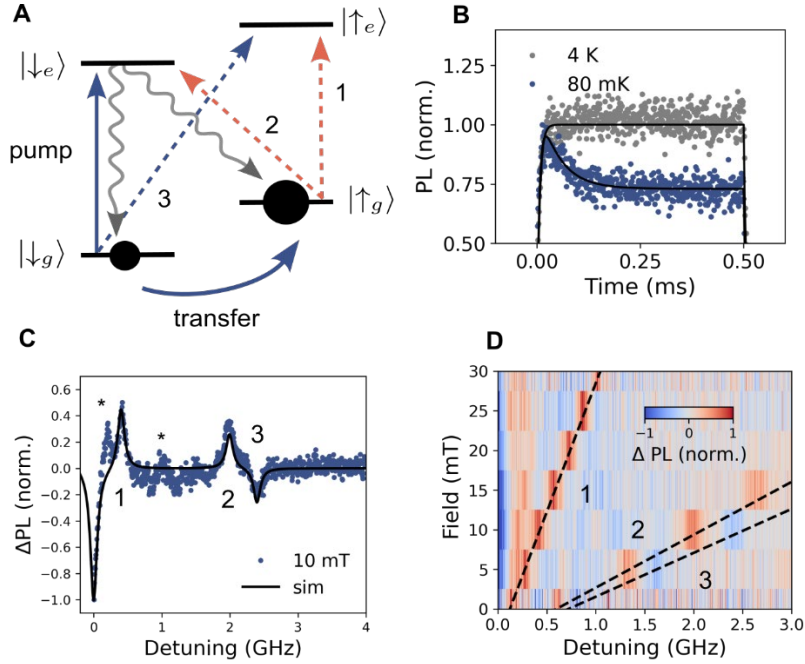

**Fig. 5. Ground-state spin pumping at sub-Kelvin temperature.** (**A**) Spin level diagram for optical spin pumping via spin-selective optical transition (blue) and non-spin-selective decay (grey) in the limit of slow spin relaxation relative to spin-non-conserving excited-state decay. Dashed lines indicate remaining optical transitions with color signifying the effect on emission (red for enhanced and blue for depleted emission). (**B**) Observed time-dependent drop in photoluminescence under resonant excitation at $T_{mxc} = 80$ mK (blue) and comparison to $T_{mxc} = 4$ K (grey) for bulk crystal (50ppm erbium doping) where no depletion is observed for zero applied magnetic field. (**C**) Differential photoluminescence at 80 mK and applied magnetic field $B = 10$ mT with hole at zero detuning and side-hole marked at (3) $g_e \mu_B B/h$ and anti-holes due to two-color spin mixing at offset frequencies of (1) $|\Delta g| \mu_B B/h$, and (2) $g_g \mu_B B/h$ (simulation in black, here $\Delta PL$ uses electro-optic modulation to generate frequency detuning with subtraction of off-resonant Lorentzian background). Asterisks denote artifacts due to harmonics at half integer multiples of resonance frequencies. (**D**) Magnetic field dependence of side-holes and anti-holes is shown as dashed lines with slope determined by $g_e$ and $g_g$ values measured in Fig. 3 against experimental data to confirm spin transfer into the opposite spin state.